\documentclass[twocolumn,showpacs,preprintnumbers,amsmath,amssymb]{revtex4}
\usepackage{graphicx}% Include figure files
\usepackage{dcolumn}% Align table columns on decimal point
\usepackage{bm}% bold math

\begin{document}

\preprint{}

\title{
Coexistence of a nearly spin-singlet state and antiferromagnetic long-range order in quantum spin system Cu$_2$CdB$_2$O$_6$
}

\author{Masashi Hase}
 \email{HASE.Masashi@nims.go.jp}
\author{Masanori Kohno}
\author{Hideaki Kitazawa}
\author{Osamu Suzuki}
\author{Kiyoshi Ozawa}
\author{Giyuu Kido}
\author{Motoharu Imai}
\author{Xiao Hu}

\affiliation{%
National Institute for Materials Science (NIMS), 1-2-1 Sengen, Tsukuba 305-0047, Japan
}%

\date{\today}% It is always \today, today,

\begin{abstract}

A 1/2 magnetization plateau in magnetic fields above 23 T and antiferromagnetic (AF) long-range order (AFLRO) in low fields were found in Cu$_2$CdB$_2$O$_6$. 
Experimental results agree with quantum Monte Carlo results for an expected spin system. 
There are two kinds of Cu sites [Cu(1) and Cu(2)], which are located adjacent to each other. 
Unexpectedly, spins on the Cu(1) and Cu(2) sites are in a nearly spin-singlet state and form AFLRO, respectively, although interactions between the Cu(1) and Cu(2) spins cannot be ignored. 
Cu$_2$CdB$_2$O$_6$ is the first material which shows such coexistence in an atomic scale. 

\end{abstract}


\maketitle

Coexistence or competition of plural different states has attracted much attention in condensed matter physics. 
Examples of such coexistence are an ordered stripe phase \cite{Tranquada95} and spatial distribution of electronic energy gaps \cite{Lang02} in high-$T_{\rm c}$ cuprate superconductors, a mixture of insulating regions with charge ordering and metallic ferromagnetic domains in manganites \cite{Uehara99}, and coexistence of superconductivity and antiferromagnetic order in {\it R}Ni$_2$B$_2$C ({\it R} = Ho or Dy) \cite{Eisaki94}. 
These phenomena result from strong correlations among electrons or spins, and therefore similar phenomena can be seen in various fields of condensed matter. 

Let us now consider whether coexistence of plural different states occurs in quantum spin systems consisting of localized spins or not. 
In quantum spin systems, various states are realized such as a spin-singlet state with a spin gap and a magnetically long-range ordered state. 
The former state is caused by quantum-mechanical effects, while the latter can be usually understood in semiclassical pictures. 
These two states are representatives in quantum spin systems, and quite different from each other. 
There are examples of spin states, which possess characters of a spin-singlet state with a spin gap and antiferromagnetic (AF) long-range order (AFLRO). 
In spin-gap systems such as the spin-Peierls system CuGeO$_3$ \cite{Hase93} and the two-leg ladder system SrCu$_2$O$_3$ \cite{Dagotto92} doped with impurities \cite{Hase96,Azuma97}, the ground state shows AFLRO, and two kinds of magnetic excitations exist. 
One is a high-energy excitation corresponding to a singlet-triple gap in pure systems, and the other is a low-energy spin-wave excitation reflecting the existence of AFLRO. 
However, Cu sites of nearly spin-singlet states and those in AFLRO cannot be distinguished. 
Thus, the spin states in the spin-gap systems doped with impurities are different from the above-mentioned examples of coexistence. 
There is also an example of a magnet including both spin-singlet states and AFLRO which are almost decoupled. 
In NH$_4$CuCl$_3$, the spin system consists of three different $S=1/2$ dimer subsystems \cite{Matsumoto03}. 
States in two dimer subsystems are spin singlet, whereas AFLRO appears in another subsystem in low fields. 
Exchange interactions between spins in the spin-singlet states and those forming AFLRO are small in comparison with major interactions, meaning that the spin-singlet states and AFLRO are almost independent. 
Therefore, the coexistence of spin-singlet states and AFLRO in NH$_4$CuCl$_3$ can be understood easily. 

At present, no magnet satisfying the following situation, which cannot be easily expected, has been reported so far. 
Spins on some sites are in a nearly spin-singlet state and those on the other sites exhibit AFLRO, although exchange interactions between these two kinds of spins cannot be ignored. 
In such a case, it is natural to consider that either a spin-singlet state or AFLRO appears \cite{Masuda04}. 

Unexpectedly, we found coexistence of a nearly spin-singlet state and AFLRO in Cu$_2$CdB$_2$O$_6$. 
As is shown in Fig. 1, there are two kinds of Cu sites [Cu(1) and Cu(2)] in this cuprate \cite{Munchau95}. 
In low magnetic fields, spins on the Cu(1) sites are in a nearly spin-singlet state, and those on the Cu(2) sites form AFLRO, although interactions between the Cu(1) and Cu(2) spins cannot be ignored. 
Cu$_2$CdB$_2$O$_6$ is the first material showing such coexistence in quantum spin systems. 

We show an expected spin system of Cu$_2$CdB$_2$O$_6$ in Fig. 1. 
Taking the composition and reasonable valence of each ion (Cd$^{2+}$, B$^{3+}$, and O$^{2-}$) into account, all Cu ions are divalent and have $S = 1/2$. 
Their positions are shown schematically in Fig. 1(a). 
Two crystallographic Cu sites [Cu(1) and Cu(2)] exist there along with five kinds of short Cu-Cu bonds, which are summarized in Table I. 
From Cu-O-Cu angles, signs of exchange interactions in Cu-Cu bonds 1, 2, and 3 are presumed to be AF. 
Values of the exchange interactions are defined as $J_1$, $J_2$, and $J_3$, respectively. 
In bonds 4 and 5, no neighboring oxygen site exists to which both two Cu sites in each bond are connected. 
For that reason, exchange interactions in these bonds are expected to be smaller than $J_1$, $J_2$, and $J_3$. 
In addition, Cu-Cu distances in the other bonds are larger than 4.64 \AA. 
Consequently, Cu$_2$CdB$_2$O$_6$ is probably as the first approximation a compound including a unique spin system formed by $J_1$, $J_2$, and $J_3$ represented in Fig. 1(b). 

\begin{table*}
\caption{\label{table1}
Interatomic distances and angles in Cu$_2$CdB$_2$O$_6$. 
}
\begin{ruledtabular}
\begin{tabular}{ccccccc}
& Cu-Cu distance (\AA) & Cu-O-Cu path & angle (${}^{\circ}$) & Cu-O distance (\AA) & number of path & sign \\
\hline
bond 1 ($J_1$) & 2.98 Cu(1)-Cu(1) & Cu(1)-O(3)-Cu(1) & 99.24 & 1.96, 1.96 & 2 & AF \\
bond 2 ($J_2$) & 3.22 Cu(1)-Cu(2) & Cu(1)-O(2)-Cu(2) & 118.48 & 1.89, 1.86 & 1 & AF \\
bond 3 ($J_3$) & 3.40 Cu(2)-Cu(2) & Cu(2)-O(1)-Cu(2) & 102.39 & 1.95, 2.40 & 1 & AF \\
bond 4 & 3.40 Cu(1)-Cu(1) & none & & & & \\
bond 5 & 3.56 Cu(1)-Cu(1) & none & & & & \\
\end{tabular}
\end{ruledtabular}
\end{table*}

Crystalline powder of Cu$_2$CdB$_2$O$_6$ was synthesized using solid-state reaction method. 
We used X-ray diffraction measurement to confirm the formation of Cu$_2$CdB$_2$O$_6$ and the absence of other materials. 
Using an inductively coupled plasma atomic emission spectrometer (ICP-AES), we estimated the composition of our sample as Cu$_{1.98}$CdB$_{2.01}$O$_{6.14}$, which closely resembled Cu$_2$CdB$_2$O$_6$. 
We measured magnetizations up to 5 T using a superconducting quantum interference device (SQUID) magnetometer produced by Quantum Design. 
High-field magnetizations up to 30 T were measured using an extraction-type magnetometer in a hybrid magnet at the High Magnetic Field Center, NIMS.
Specific heat was measured using relaxation technique with PPMS, which was produced by Quantum Design. 
We calculated susceptibility and magnetization of the spin system in Fig. 1(b) by a quantum Monte Carlo (QMC) technique using the directed-loop algorithm in the path-integral formulation \cite{Syljuasen02}. 
The number of sites and Monte Carlo samples in the QMC simulations are one thousand and about one million, respectively. 
Finite-size effects and statistical errors are negligible in the scale of figures represented in this Letter. 

The solid curves in Figs. 2(a) and 2(b) represent temperature ($T$) dependence of magnetic susceptibility $\chi (T)$ of Cu$_2$CdB$_2$O$_6$ in the magnetic field of $H = 0.1$ T. 
The susceptibility at high $T$ obeys the Curie-Weiss law, which is expressed as $C/(T+ \theta)$.
Here $C$ and $\theta$ are a Curie constant and Weiss temperature, respectively. 
The inset of Fig. 2(a) shows inverse susceptibility. 
We roughly estimated $C$ as 0.448 emu K/Cu mol from a slope of the inverse susceptibility above 180 K and $\theta$ as 47.4 K from the intersection of the dashed line and the $T$ axis. 
We determined independently the powder-averaged gyromagnetic ratio of Cu$^{2+}$ as $g = 2.10$ in an electron spin resonance (ESR) measurement at room temperature. 
Thereby, a value for $C$ was calculated as $C=0.414$ emu K/Cu mol assuming that all the Cu ions have $S=1/2$. 
The two $C$ values are similar to each other. 
This result indicates that almost all the Cu ions have $S=1/2$. 

The susceptibility of Cu$_2$CdB$_2$O$_6$ has a maximum around 11 K and seems to reach a finite value at 0 K. 
The inset of Fig. 2(b) represents $d \chi (T)/dT$. 
A lambda-type behavior is observed around 10 K. 
Figure 2(c) shows specific heat $C(T)$ in 0 T, which has a peak at 9.8 K. 
Taking the expected signs of the exchange interactions into account, AFLRO probably appears at $T_{\rm N}=9.8$ K. 
Besides, we found a spin-flop transition around 1.5 T in magnetization $M(H)$ measured by the SQUID magnetometer at 2.9 K. 
A small value of a spin-flop field $H_{\rm SF} \sim 1.5$ T is reasonable because magnetic anisotropy of spins on Cu$^{2+}$ ions is usually small. 
For example, $H_{\rm SF}$ is about 1 T in Cu$_{0.96}$Zn$_{0.04}$GeO$_3$ \cite{Hase95}. 
The dotted curve in Fig. 2(a) shows the QMC result of $\chi (T)$ for the spin system in Fig. 1(b). 
It is explained later. 

The solid curve in Fig. 3 represents $M(H)$ at 2.9 K. 
The most prominent feature is a 1/2 magnetization plateau above 23 T. 
Diamonds represent the QMC result of $M(H)$ at 2.9 K for the spin system in Fig. 1(b) with $J_1 = 160$, $J_2 = 38.8$, and $J_3 = 9.7$ K and can reproduce well the experimental $M(H)$. 
The QMC result of $\chi (T)$ for the same model is indicated by the dotted curve in Fig. 2(a) and agrees with the experimental $\chi (T)$. 
Therefore, the spin system in Fig. 1(b) can explain the magnetic properties of Cu$_2$CdB$_2$O$_6$ at least as the first approximation. 
As mentioned above, however, AFLRO appears at low $T$, which is caused by weak three-dimensional (3D) couplings ignored in the spin model. 
The discrepancy between experimental $\chi (T)$ and the QMC result of $\chi (T)$ at low $T$ is due to the appearance of AFLRO.

Spin states can be understood as follows (Fig. 4). 
The appearance of the 1/2 magnetization plateau means that half of spins are in a nearly spin-singlet state with a dimer gap, which corresponds to the magnetic field to break the nearly spin-singlet state, and that the other spins are almost polarized parallel to the applied field. 
Therefore, the magnetization cannot increase in the plateau region with an increase in the magnetic field. 
The QMC results in Fig. 3 show that moments of the Cu(1) spins (open circles) are very small in the low-field regime, indicating that the Cu(1) spins form nearly singlet pairs with the dimer gap below an end field of the plateau. 
The inset of Fig. 3 shows the QMC results of $M(H)$ for a wider region of $H$. 
The magnetization starts to increase again around 108 T, which means that a value of the dimer gap is about 152 K for $g = 2.1$. 
The dimer gap corresponds to binding energy of singlet pairs and is mainly determined by the $J_1$ interaction. 
Thus, the value of the dimer gap (152 K) is close to $J_1$ (160 K). 
The QMC results in Fig. 3 show that moments of the Cu(2) spins (open squares) are almost polarized in the plateau region. 
Since the moments of the Cu(1) spins are small below 108 T, an effective interaction between two Cu(2) spins through the AF dimer formed by the Cu(1) spins is small. 
Thus, the Cu(2) spins can be almost saturated above 23 T in spite of the large values of $J_1$ and $J_2$ [Fig. 4(b)]. 
AFLRO can appear in the Cu(2) spins [Fig. 4(a)] by this effective interaction, the $J_3$ interaction, and other weak 3D couplings which are omitted in our model \cite{Comment1}. 
The distance of the nearest-neighbor Cu(1)-Cu(2) bond (bond 2) is only 3.22 \AA \ and the interaction in this bond (38.8 K) cannot be ignored in comparison with other interactions. 
Nevertheless, the coexistence of the nearly-singlet state of the Cu(1) spins and AFLRO of the Cu(2) spins appears. 
We can say that this is "coexistence in an atomic scale" and such coexistence in quantum spin systems has not been found before the present study. 

As was described, both the Cu(1) and Cu(2) spins have non-zero moments. 
The average value of the total spins is, on the contrary, exactly $S = 0.25$ in the plateau region 
This result indicates that a simple idea, in which some spins are perfectly spin singlet and others are fully polarized, cannot explain the magnetization plateau in Cu$_2$CdB$_2$O$_6$. 
Similar spin states must appear in other spin systems possessing magnetization plateaus. 
It is proved that magnetization curves at 0 K may have plateaus at $m'$ satisfying the formula of $n(S'-m')=$ integer \cite{Oshikawa97}. 
Here, $n$ is period of ground state, and $S'$ and $m'$ are the total spin and magnetization per unit cell, respectively. 
In our model, $n=1$, $S'=2$ (four $S=1/2$ spins), and the plateau appears at $m'=1$. 
Thus, our result is consistent with the theorem of Ref. [\onlinecite{Oshikawa97}]. 
We also calculated magnetization curves at 5.5 and 15 K and confirmed that the values of magnetization at the plateau are also $S = 0.25$. 

In conclusion, we observed the 1/2 magnetization plateau in magnetic fields above 23 T and the antiferromagnetic (AF) long-range order (AFLRO) in low magnetic fields in Cu$_2$CdB$_2$O$_6$. 
The experimental results explained herein are consistent with the quantum Monte Carlo results for the expected spin system.
The two kinds of Cu sites [Cu(1) and Cu(2)] are located adjacent to each other.
The spins on the Cu(1) sites are in the nearly spin-singlet state, and a finite energy (dimer gap) is necessary to break the nearly spin-singlet state. 
The spins on the Cu(2) sites are almost polarized in the 1/2 plateau region, whereas they form AFLRO in low fields.
Thus, unexpectedly, coexistence of the nearly spin-singlet state and AFLRO appears, although the interaction between the Cu(1) and Cu(2) spins cannot be ignored. 
To our knowledge, Cu$_2$CdB$_2$O$_6$ is the first material in quantum spin systems which shows such coexistence in an atomic scale. 
Future studies must address nuclear magnetic resonance (NMR) measurements and theoretical determination of a phase diagram in the spin system. 

\begin{acknowledgments}

We are grateful to K. Uchinokura, A. Tanaka, and N. Tsujii for invaluable discussion, to K. Yamada and H. Yamaguchi for ICP-AES measurements, to H. Yamazaki for ESR measurements, to M. Kaise for X-ray diffraction measurements, and to H. Mamiya and T. Furubayashi for use of their PPMS machine. 
This work was supported by grants for basic research from NIMS and by a Grant-in-Aid for Scientific Research from the Ministry of Education, Culture, Sports, Science, and Technology. 

\end{acknowledgments}

\newpage %Just because of unusual number of tables stacked at end

\begin{references}
\bibitem{Comment1}
In the spin system of Fig. 1(b) without other interactions, the ground state must be intrinsically a spin-singlet state with a spin gap in the zero field for $J_2 > 0$. When the $J_1$ interaction is large compared with the others, the moments of the Cu(1) spins are reduced in low magnetic fields as is shown in Fig. 3. Thus, the effective interaction between two Cu(2) spins through the AF dimer formed by the Cu(1) spins is small. As a result, the spin gap is expected to be small, because it vanishes when the effective interaction is zero. In the parameter regime corresponding to Cu$_2$CdB$_2$O$_6$, the spin gap is too small to be detected in the scale of Figs. 2(a) and 3. Thus, it is expected that the spin gap is easily destroyed by weak 3D couplings.
\end{references}

\begin{figure} 
\caption{
(color). (a) 
Schematic drawing of Cu$^{2+}$-ion positions in Cu$_2$CdB$_2$O$_6$. 
Red and blue circles indicate Cu(1) and Cu(2) sites, respectively. 
Red, green, blue, dashed black, and solid black bars represent Cu-Cu bonds 1, 2, 3, 4, and 5, respectively. 
(b) 
An illustration of the spin system in Cu$_2$CdB$_2$O$_6$. 
The Hamiltonian is expressed as ${\cal H} = \sum_{i} [J_1 S_{i,2} \cdot S_{i,3} + J_2 (S_{i,1} \cdot S_{i,2} + S_{i,3} \cdot S_{i,4}) + J_3 (S_{i,1} \cdot S_{i+1,1} + S_{i,4} \cdot S_{i+1,4})]$. 
Exchange interactions in the bonds 1, 2, and 3 ($J_1$, $J_2$, and $J_3$, respectively) are AF and mainly determine magnetic properties. 
}
\label{Fig. 1}
\end{figure}

\begin{figure} 
\caption{
Magnetic susceptibility $\chi (T)$ below 300 K in (a) and 50 K in (b). 
The solid and dotted curves respectively indicate $\chi (T)$ of Cu$_2$CdB$_2$O$_6$ and the QMC result of $\chi (T)$ for the spin system in Fig. 1(b) with $J_1 = 160$, $J_2 = 38.8$, and $J_3 = 9.7$ K. 
The insets in (a) and (b) show respectively inverse $\chi (T)$ and $d \chi (T)/dT$ curves. 
The dashed line in the inset of (a) denotes the fitted line to inverse $\chi (T)$ above 180 K. 
A single division of the vertical scale in the inset of (b) means $5 \times 10^{-4}$ emu/Cu mol K. 
(c) Specific heat $C(T)$ of Cu$_2$CdB$_2$O$_6$ below 50 K. 
The inset shows $C(T)$ below 300 K.
}
\label{Fig. 2}
\end{figure}

\begin{figure} 
\caption{
Magnetization of Cu$_2$CdB$_2$O$_6$ (solid curve) expressed as a spin value $S = M(H)/g \mu_{\rm B} N_{\rm A}$ where $\mu_{\rm B}$ and $N_{\rm A}$ are the Bohr magneton and Avogadro's constant, respectively. 
Diamonds, circles, and squares respectively indicate the QMC results of $M(H)$ for total, Cu(1), and Cu(2) spins of the spin system in Fig. 1(b) with $J_1 = 160$, $J_2 = 38.8$, and $J_3 = 9.7$ K. Curves 1, 2, and 3 in the inset shows the QMC results of $M(H)$ up to 140 T for total, Cu(1), and Cu(2) spins, respectively.
}
\label{Fig. 3}
\end{figure}

\begin{figure} 
\caption{
(a) Schematic picture of the spin state in low fields. 
Ellipses suggest that the spins on Cu(1) sites are in a nearly spin-singlet state with a dimer gap. 
Solid arrows indicate that the Cu(2) spins form AFLRO, but the direction of the ordered spins has not been determined experimentally. 
(b) The spin state in the 1/2 magnetization plateau region. 
Solid arrows indicate that the spins on Cu(2) sites are almost polarized parallel to the applied field. 
}
\label{Fig. 4}
\end{figure}

\end{document}